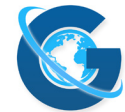

**Review Article**

# Determination of Canine Long Bone Ultimate Tensile Strain by Digital Image Correlation

## Beatrice Böhme[1], Cédric Laurent[2], Olivier Milis[3], Jean-Phillippe Ponthot[4], Marc Balligand[1*]

[1]Department of Clinical Sciences, Small Animal Surgery Service, Faculty of Veterinary Medicine, University of Liège, Liège, Belgium

[2]ENSEM - LEMTA, University of Nancy, Nancy, France

[3]Electromechanics, Department of Physics, Faculty of Sciences, University of Liège, Liège, Belgium

[4]Department of Aerospace and Mechanics, University of Liège, Liège, Belgium

**\*Corresponding author:** Marc Balligand, Department of Clinical Sciences, Small Animal Surgery Service, Faculty of Veterinary Medicine, University of Liège, Sart Tilman, 4000 Liège, Belgium

## Abstract

Creating finite element models for simulation of bone behaviour, fracture occurrence and propagation requires feeding the system with adequate mechanical data. To date, little is known about the mechanical behaviour of long canine bones, and in cases of modelling a bone-breakage scenario the ultimate strain at failure still needs to be determined. Extrapolation from human literature or other species is uninteresting as differences of mechanical properties between species are expected (Vahey et al, 1987) [1]. Our purpose was to measure the ultimate tensile strain of canine long bones by Digital Image Correlation (DIC), an optical technique to measure strain under load and a promising measurement method for our purpose. Tension tests of cortical bone strips and flexion tests of entire bones were performed, and the ultimate tensile strain was measured and compared between different specimen and testing scenarios.

## Introduction

The mechanical behaviour of long canine bones is an important topic in cases of decision-making for fracture repair and the establishment of computational finite-element models. The behaviour of canine bone under loading has already been described by load-displacement curves from mechanical testing [2,3], but the ultimate tensile strain of canine cortical bone still needs to be precisely determined. Traditionally, strain gauges have been used to measure surface strain and deformation, measuring displacement between two surface points. But this technique has its limitations: as strain gauges can only provide data locally, they might not be placed at the initial fracture site and are quite difficult to hold in place when glued onto an irregular organic surface. Digital Image Correlation (DIC) was more suitable for our purposes, an optical technique which has already been validated to measure deformation and strain under load [4-9] for organic inhomogenous anisotropic non-linear composite materials like bone [4,10,11]. This technique provides an overall picture of the deformations affecting a large surface.

DIC is an optical non-contact method that allows for visualisation of a larger field, up to the entire bone, in a 3D configuration. DIC uses image acquisition by at least two high-speed cameras to trace displacement by comparing digital images before and after deformation. To allow tracing of image pixels, a random speckle pattern is created on the surface by manually spraying a black and white paint on the specimen before testing. The change of position by displacement of points in an area of interest is processed using specific DIC software [4] and can be presented in the form of strain maps [4,8,11].

## Materials and Methods

### Specimen

Four pairs of canine fore- and hind-limbs were harvested from adult dogs that weighed 20-30 kg, were between 1.5-8 years of age, and had been euthanised for reasons unrelated to this study.

Humeri, radii and femurs were isolated, soft tissues were removed, and samples were wrapped in saline-soaked sponges and stored at minus 20°C. Specimen for mechanical testing were thawed at room temperature overnight and kept moistened.

**Preparation of Samples for Tensile Testing**

Humeral diaphysis (n=4) were cut in a craniocaudal direction creating equal halves. The flat medial part was trimmed down to a dumbbell-shaped diaphyseal specimen of 1cm x 6cm with a thinned part at the central area. The large ends were plotted distally and proximally in polyester resin blocs (Motip® Wolvega, The Netherlands) and placed in custom-made conical press fit cylinders (diameter 50.00mm-32.00mm, height 40.00mm). Specimens were coloured with DIC-specific colours (white: Mop 04036; black: Mop 04031) and centred in the electrohydraulic testing machine (Zwick, 100kN maximum load). The cylinders were fixed to a rotulated jig of the testing machine (see Figure 1). Testing was performed at a constant displacement of 0.01mm/s until breakage.

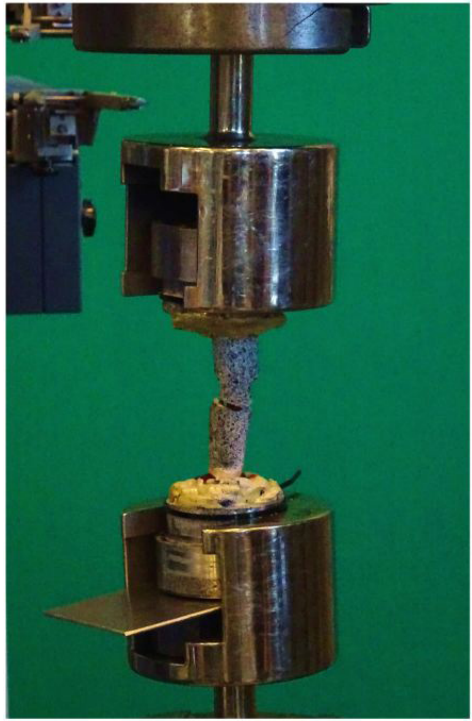

**Figure 1:** Specimen after testing, placed in the testing machine

**Preparation of Samples for Flexion Testing**

Canine radii and femurs were left intact and coloured with DIC-specific colours (white: Mop 04036; black: Mop 04031). Specimens were centred in the electrohydraulic testing machine (Zwick, 100kN maximum load). Testing was performed at a constant displacement of 0.2mm/s (whole radius, flexion testing) and 0.1mm/s (whole femur, flexion testing). In addition to the drop in stiffness, acoustic emission recording was performed during whole-femur testing to confirm the moment of crack and drop in stiffness (acoustic emission sensors Micro-80 and a PCI2 Mistras system, Euro Physical Acoustics, Software AEWin).

**Digital Image Correlation (DIC)**

Two high-speed cameras (AVT Pike, type f032B, Kodak KAI340; resolution 640 x 480 pixels), connected to a PC unit, were placed in front of the specimen at different angles, allowing an overlapping image view. Digital image acquisition and testing start were synchronised. Images and videos were recorded at a frequency of 5-100 Hrtz. Images were acquired for 3D treatment by Vic Snap 2007 software (Correlated Solutions); synchronisation and image correlation were performed using Vic3D software (Correlated Solutions). Analysis was performed in a 170mm zone at a resolution of 9 x 9 pixels (radius, humerus) and 11 x 11 pixels (femur). Nine representative points of maximal tensile strain were selected at maximal load just before breakage (see Figure 2). Data obtained from tension and flexion tests were compared.

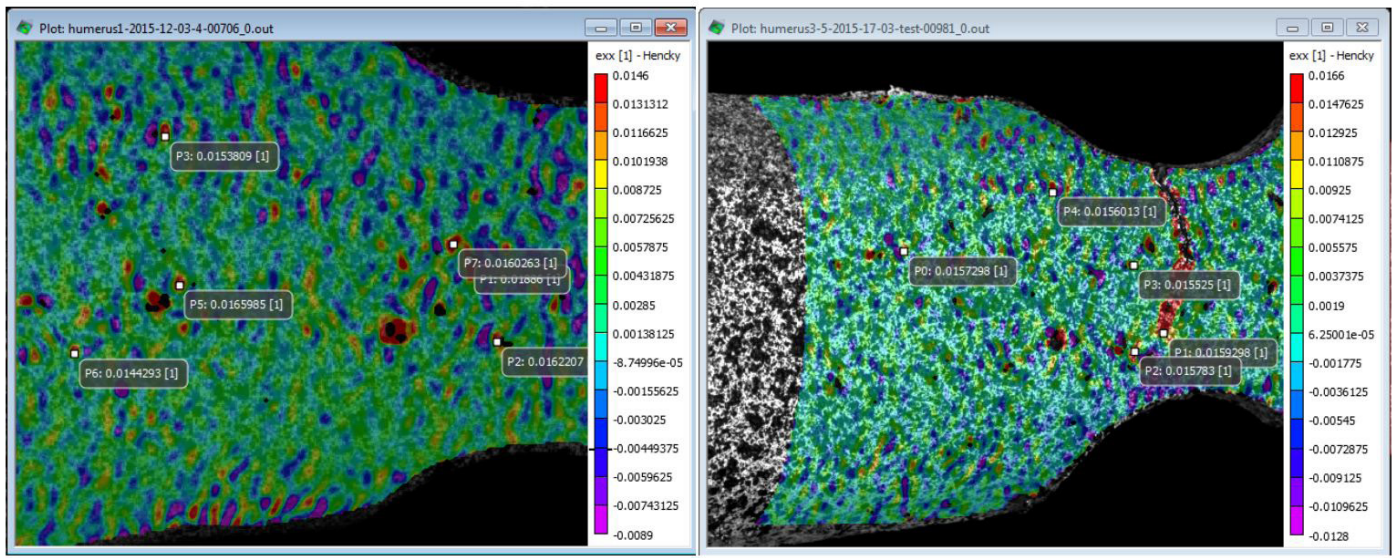

**Figure 2:** Deformation plot humerus before and during breakage.

### Statistical Analysis

Statistical analysis was performed using a linear model with two factors: factor one being the bone type (humerus, radius and femur); and factor two being the individual bone (n=4 per bone type, nested within its type). The parameters have been estimated using the SAS (v9.4) GLM procedure.

### Results

All specimens fractured through a sudden brittle crack at the tension side within the thinned part (humeri) or diaphysis (radii and femurs). The crack was barely observable, but bone rupture was attested by a sudden drop of the specimen stiffness. Acoustic emission recording during whole-femur testing confirmed the moment of crack and drop in stiffness. Ultimate tensile strain was 1.745% ± 0.091 for humeri tested in tension, 1.724% ± 0.079 for femora and 1.718% ± 0.079 for radii, both tested in flexion. Mean ultimate tensile strain for all bones tested was 1.73% ± 0.083 (see Table 1, Figure 3). There was no statistical difference in ultimate tensile strain between dumbbell-shaped specimen (humerus) tested in tension and whole bones (radius, femur) tested in flexion. Differences between bone types have been assessed and no significance was obtained at the 5% threshold.

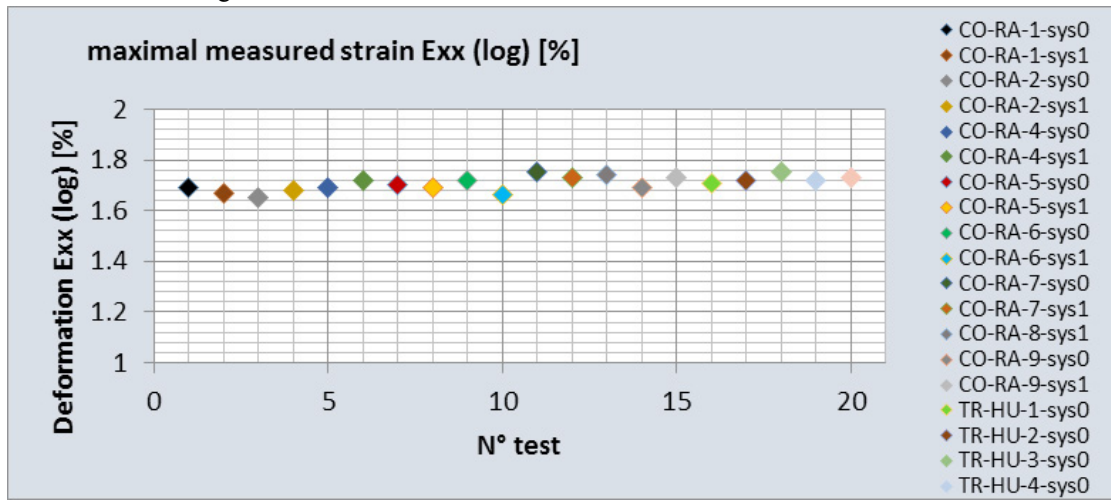

**Table 1:** Ultimate tensile strain values for humerus strips tested in tension, and for whole radii and femurs tested in flexion.

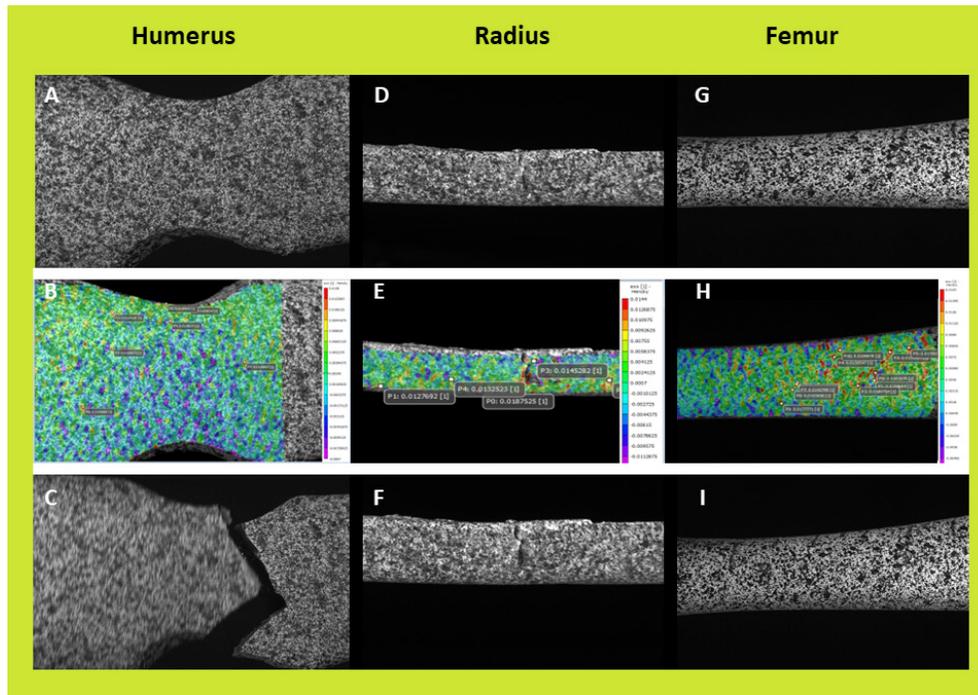

**Figure 3:** Examples of high-resolution images just before (A, D, G) and after bone breakage (C, F, I), and the corresponding strain map (B,E,H) of humerus, radius and femur. Resolution: 9 [px] X 9 [px] (A-F) and 11 [px] x 11[px] (G-I).

## Discussion

Failure of cortical bone is guided by a critical level of strain, the 'ultimate tensile strain', which determines the maximal elongation before breakage occurs. Bone as a composite material is made of organic cells (osteoblasts, osteoclasts, osteocytes etc.) and extracellular matrix. Extracellular matrix contains collagenic fibrils organised in collagen fibres and anorganic hydroxyapatite material (mineral structure) [12]. In cortical bone, collagenic fibres are oriented in concentric cylinders around the Haversian canals (vessels), forming so-called osteons. Whereas fibres are mainly responsible for bone elasticity, incorporated minerals determine compressive strength and resistance to failure [13]. Available data suggest that the structure of bone seem to have greater influence on its tensile properties and stiffness than density [14-16]. Bone breakage may develop due to tensile failure of collagen fibrils or debonding of organic matrix caused by shear failure and transverse separation of fibrils from each other [17,18]. Some mechanical features, like yield and ultimate strain, seem to be independent of trabecular orientation in the specimen [1]. The theory that ultimate failure in bone results from failure of the organic matrix has been supported [19]; bone mineral density might not be predictive [15,16].

At time of bone failure, strain seems to concentrate at the crack's tip and around the osteocyte lacunae [20]. Microscopic strain patterns are highly heterogeneous and in some locations are similar to the observed microdamage around osteocyte lacuna, indicating that the resulting strains may represent the precursors to microdamage [20]. Failure occurs when organic material reaches the ultimate tensile strain and a fracture line propagates from microscopic to macroscopic level. This occurs extremely quickly and cannot be captured by the human eye or even high-speed cameras. Load-displacement curves and acoustic emission recording and its correlation can help to distinguish the moment of failure. High-speed cameras with even higher frequency of recording were not available for the study. We used the acoustic emission recording to verify that bone is a composite material and to determine the moment of failure in femoral whole-bone specimens where no crack was visible in DIC. DIC was the method of choice to measure strain in this study. Instead of gluing extensometers on the bone surface or simple extrapolation from the load-displacement curve, DIC allowed us to measure the ultimate tensile strain with visualisation of the entire bone in a three-dimensional configuration. DIC is an accepted and suitable method for measurements of ultimate tensile strain in composite materials and has already been shown to measure strain on bone surfaces [8,9,11,21,22]. Image acquisition is performed from at least two high-speed cameras to trace displacement of surface points. Comparing the localisation of specific surface points before and after deformation allows us to quantify the change of position by displacement of points in an area of interest. Strain pattern can be visualised by specific DIC software [4] and can be presented in the form of strain maps [4,8,11] (see Figures 2 and 4).

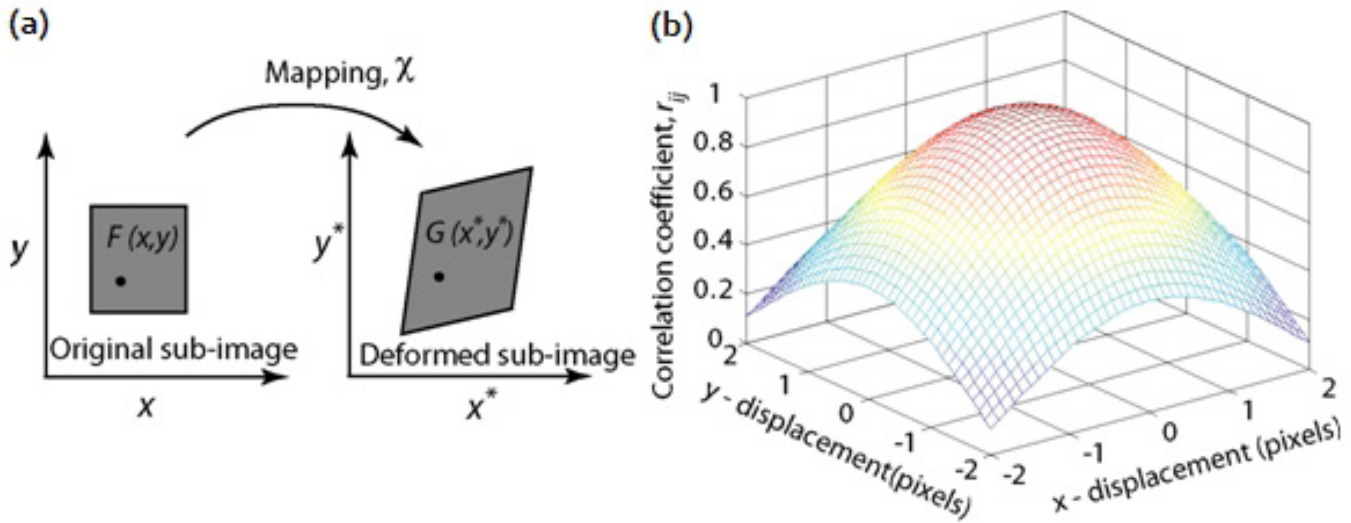

**Figure 4:** Strain map showing displacement of surface points during loading (source: Public Domain, https://en.wikipedia.org/w/index.php?curid=14642925)

Mean ultimate tensile strain for medial humerus, whole radius and whole femur in adult dogs (1.5-8y) was 1.73% ± 0.083. There was no statistical difference between bone type or methodology (specimen in tension or whole bone in flexion). These data are comparable to already published data in the form of calculations [23] and measurements [15,24] of cortical bone ultimate tensile strain, and to data obtained in younger men [14]. The variation in ultimate tensile strain, from 1.07% (femur of older men) [14] to 2.57% (bovine) [15], might be due to variation in species [1], bone type [14,25], donor age [14], diaphyseal sampling or bone morphology [23,26,27], pretreatment of samples [14,28], and variation in testing methods. Results for mechanical testing of organic material may be influenced by the requirements of storage prior to testing, especially with multiple freeze-thaw cycles and dehydration. Our results should not be influenced by this effect, as various authors have confirmed that cortical specimen undergoing a single freeze-thaw cycle (at minus 20°C storage) do not show significant changes in mechanical properties [21,29-32].

To address the issue of bone morphology, three different types of bones with various shapes (radius, humerus and femur) were used in our study. On a macroscopic level, there was no statistical difference in ultimate tensile strain results between these three bones. The tension tests performed using dumbbell-shaped specimen of the humerus, chosen according to Feng's work [33] did not deliver statistically different results in ultimate tensile strain. This was considered confirmation of realistic measurements obtained in this study. Nine representative points of maximal tensile strain within or close to the fracture line were selected at maximal load before breakage on the bone surface (see Figure 2). Analysis in our study was performed at a resolution of 9 x 9 pixels (radius, humerus) and 11 x 11 pixels (femur) at the macroscopic level of bone surface, representing a zone on the bone surface of 2.4 x 2.4mm and 2.93 x 2.93mm, respectively. Nine points of highest strain values were selected as representative strain locations just before the moment of failure. Bone breakage is extremely quick, and even high-speed cameras can hardly capture occurrence of the fracture line. Therefore, the highest measurable strain can be considered to represent the ultimate tensile strain. The ultimate strain across the fracture line and occurrence of the fracture line were confluent, but detectable by a drop in the load-displacement curve and an audible crack on acoustic emission recording.

Strain is not evenly distributed within the specimen and depends on the resolution at which it is looked: Nicolella et al. used DIC to compare strain propagation from the macroscopic to the microscopic level and found that experimentally-determined macroscopic strains of approximately 0.2% reach levels of over 3% at the osteocyte lacuna at bone matrix level, which means a fifteen-times increase in the applied macroscopic strain [20]. When ultimate strain locations at the microscopic level start to fuse, they create a weak point for fracture initiation and initiate the fracture propagation up to macroscopic level along the weakest composites with the highest strain [20]. To be able to compare macroscopic surface measurements and data already available, we chose a representative resolution for our surface measurements. To our knowledge, this is the first study to measure ultimate tensile strain of canine diaphyseal cortical bone tissue by DIC. DIC is a suitable method to measure ultimate tensile strain in canine bone. In further studies, including ultimate tensile strain in finite element models will be useful to create a realistic scenario of canine long bone behaviour and to explore fracture pattern and osteosynthetic repair behaviour [3]. This may answer questions that are still not answered. How do fractures propagate through bone? What is the value-added of a specific osteosynthetic repair by surgeons? Which elasticity is allowed for the implant to speed up healing without risk of bone or implant failure? Is there an option to develop implants more suitable during the stabilisation period? Finite element models might answer these questions, but they are only reliable if they are constructed with adequate mechanical data.

## Conclusion

The purpose of this study was to determine the ultimate tensile strain of canine cortical bone tissue by Digital Image Correlation (DIC). Based on our work and others [22], DIC is a suitable and valuable method to measure ultimate tensile strain in canine long bones. Mean ultimate tensile strain for canine bone diaphysis was 1.73%, which is comparable to values obtained in younger men [14]. There was no statistical difference of ultimate tensile strain between tests in tension (humerus strips) and flexion (whole radius and femur). Including the ultimate tensile strain in finite element models will be useful to create a realistic scenario of canine long bone behaviour and to explore fracture pattern and osteosynthetic repair behaviour in further studies [3].